\documentclass[a4paper,11pt]{article}

\usepackage{hyperref}
\hypersetup{
colorlinks=true,
linkcolor=magenta,
citecolor=blue,
urlcolor=black,
bookmarksnumbered=true,
bookmarkstype=toc,
bookmarksopen=false,
pdftitle={BPS Monopole in the Space of Boundary Conditions},
pdfauthor={Satoshi Ohya},
pdfsubject={non-Abelian Berry's connection},
pdfkeywords={boundary condition; non-Abelian geometric phase; 't Hooft-Polyakov monopole; supersymmetric quantum mechanics}}

\usepackage[margin=1in]{geometry}

\usepackage[adobe-utopia,cal=cmcal]{mathdesign}
\usepackage[T1]{fontenc}
\usepackage[misc]{ifsym}

\usepackage{xcolor}

\usepackage{amsmath,amsbsy,bm}

\usepackage{epic,eepic}

\usepackage[hang,font=small,labelfont=bf]{caption}
\usepackage{subfigure}

\makeatletter

\@addtoreset{equation}{section}
\makeatother

\usepackage{titlesec}
\titleformat{\section}[block]{\filright\bfseries\mathversion{bold}}{\thesection.}{0.5em}{}[\titlerule]
\titleformat{\subsection}[block]{\filright\bfseries\mathversion{bold}}{\thesubsection.}{0.5em}{}

\usepackage{cite}

\begin{document}

\vspace*{2em}
\begin{center}
{\bfseries\large BPS Monopole in the Space of Boundary Conditions}
\par\vspace*{2em}
{\normalsize Satoshi Ohya}
\par\vspace*{1em}
{\itshape\small
Institute of Quantum Science, Nihon University\\
Kanda-Surugadai 1-8-14, Chiyoda, Tokyo 101-8308, Japan}
\par\vspace{1em}
{\ttfamily\small \raisebox{-1pt}{\Letter}~\href{mailto:ohya@phys.cst.nihon-u.ac.jp}{ohya@phys.cst.nihon-u.ac.jp}}
\par\vspace{1em}
{\small (Dated: \today)}
\end{center}

\begin{abstract}
The space of all possible boundary conditions that respect self-adjointness of Hamiltonian operator is known to be given by the group manifold $U(2)$ in one-dimensional quantum mechanics.
In this paper we study non-Abelian Berry's connections in the space of boundary conditions in a simple quantum mechanical system.
We consider a system for a free spinless particle on a circle with two point-like interactions described by the $U(2) \times U(2)$ family of boundary conditions.
We show that, for a certain $SU(2) \subset U(2) \times U(2)$ subfamily of boundary conditions, all the energy levels become doubly-degenerate thanks to the so-called higher-derivative supersymmetry, and non-Abelian Berry's connection in the ground-state sector is given by the Bogomolny-Prasad-Sommerfield (BPS) monopole of $SU(2)$ Yang-Mills-Higgs theory.
We also show that, in the ground-state sector of this quantum mechanical model, matrix elements of position operator give the adjoint Higgs field that satisfies the BPS equation.
It is also discussed that Berry's connections in the excited-state sectors are given by non-BPS 't Hooft-Polyakov monopoles.
\end{abstract}

\begingroup
\hypersetup{linkcolor=black}
\tableofcontents
\endgroup

\newpage
\section{Introduction} \label{sec:1}
\begin{figure}[htb]
\centerline{\input{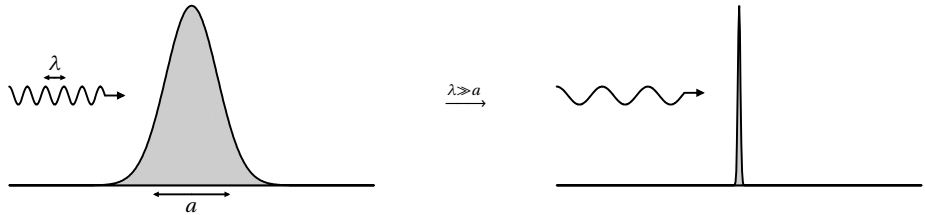}}
\caption{Long-wavelength limit. A slowly-moving particle whose de Broglie wavelength $\lambda$ is much longer than the size $a$ of the potential cannot resolve the structure of short-range interaction; that is, any short-range interaction would be effectively described by a structureless point-like interaction, or boundary condition, in the long-wavelength (i.e. low-energy) regime $\lambda \gg a$.}
\label{fig:1}
\end{figure}

Since the work of F\"{u}l\"{o}p and Tsutsui in the late 1990s \cite{Fulop:1999pf}, it has been clearly recognized that, in one-dimensional quantum mechanics, the space of all possible boundary conditions which respect self-adjointness of Hamiltonian operator (or unitarity of time-evolution) is given by the group manifold $U(2)$.
Such $U(2)$ manifold can be regarded as the \textit{theory space} of one-dimensional one-particle quantum mechanics with a single point-like interaction, because point-like interactions are all described by boundary conditions just as in the case of Dirac's delta-function potential.
Note that since the delta-function potential is characterized by a single real coupling constant, it just corresponds to a one-dimensional subspace of the full theory space $U(2)$.
Though the full $U(2)$ family of point-like interactions has not yet been realized in laboratory experiments, it is of great significance for future nanotechnology to understand the physics of point-like interactions/boundary conditions, because any short-range interaction could be approximated by a structureless contact interaction in the long-wavelength limit of incident particles; see figure \ref{fig:1}.
In other words, any short-range interactions fall onto a point of the manifold $U(2)$ as we flow to the infrared.\footnote{RG flow and universality classes for the $U(2)$ family of point-like interactions/boundary conditions are clarified in \cite{Ohya:2010zm}.}
It is therefore important for low-energy physics to investigate the $U(2)$ family of boundary conditions from various perspectives.

The purpose of the present paper is to study non-Abelian Berry's connections \cite{Wilczek:1984dh} in the space of boundary conditions, by considering a situation where boundary condition parameters are time-dependent and change adiabatically along a closed loop on the space of boundary conditions.
Such situation will be realized if the $U(2)$ family of point-like interactions is all experimentally controllable.
In the previous work \cite{Ohya:2014ska} we studied one-particle quantum mechanics on a circle with two point-like interactions at antipodal points, whose full parameter space is the direct product $U(2) \times U(2)$, and showed that non-Abelian Berry's connections in the $\mathscr{N} = 2$ supersymmetric subspace $\mathcal{M}_{\text{SUSY}} = U(1) \times U(2)/((U(1) \times U(1)) \subset U(2) \times U(2)$ are given by the Wu-Yang-like monopoles of pure $SU(2)$ Yang-Mills theory.
In this paper we would like to focus on the same setup yet with a different subspace of boundary conditions.
The main goal of this paper is to show that, for a certain $SU(2) \subset U(2) \times U(2)$ subfamily of boundary conditions, non-Abelian Berry's connection in the ground-state sector is given by the Bogomolny-Prasad-Sommerfield (BPS) monopole of $SU(2)$ Yang-Mills-Higgs theory \cite{Prasad:1975kr,Bogomolny:1975de}.
It should be noted here that the BPS monopole has already been realized as Berry's connection by Sonner and Tong in 2008 \cite{Sonner:2008be}.
They considered a quantum mechanical system for a spin-1/2 on 2-sphere with a background magnetic field and showed that, in the presence of a suitable potential, Berry's connection in the ground-state sector becomes the BPS monopole.
In this paper we would like to focus on a much simpler system; that is, a free spinless particle on a circle.
We shall show that, in the presence of suitable point-like interactions, energy spectrum for a free particle on a circle exhibits two-fold degeneracy and Berry's connection in the ground-state sector becomes the BPS monopole.

The rest of the paper is organized as follows.
In section \ref{sec:2} we set up our free-particle model in which all the energy levels become doubly-degenerate.
We solve the time-independent Schr\"{o}dinger equation exactly and see that doubly-degenerate ground states are given by zero-modes of certain first-order differential operators.
It is also discussed that the double degeneracy of energy levels results from the \textit{higher-derivative supersymmetry}, which is a nonlinear extension of $\mathscr{N} = 2$ supersymmetry introduced by Andrianov \textit{et al.} \cite{Andrianov:1993md,Andrianov:1994aj}.\footnote{Higher-derivative supersymmetry is also referred to as polynomial supersymmetry \cite{Andrianov:1994mk}, $\mathcal{N}$-fold supersymmetry \cite{Aoyama:1998nt} and nonlinear supersymmetry \cite{Plyushchay:1999qz}.}
Unlike the standard Witten model of supersymmetric quantum mechanics, higher-derivative supersymmetric models are known to be able to enjoy doubly-degenerate supersymmetric ground states \cite{Andrianov:1993md,Andrianov:1994aj}.
We then consider a time-dependent situation in which boundary condition parameters evolve adiabatically.
We first study non-Abelian geometric phase in the ground-state sector in section \ref{sec:3} and show that Berry's connection is given by the BPS monopole of four-dimensional $SU(2)$ Yang-Mills-Higgs theory.
We also show that the matrix elements of position operator give the adjoint Higgs field that satisfies the BPS equations.
In section \ref{sec:4} we explore non-Abelian geometric phases in the excited-state sectors and show that Berry's connections are given by non-BPS 't Hooft-Polyakov monopoles \cite{'tHooft:1974qc,Polyakov:1974ek}.
Conclusions are drawn in section \ref{sec:5}.
Appendix \ref{appendix:A} is devoted to computational details for singular gauge transformations that remove Dirac string singularities in Berry's connections.

Throughout the paper we will work in the units $\hbar = 2m = 1$, where $m$ is the mass of the free particle.

\section{The model: A free particle on \texorpdfstring{\mathversion{bold}$S^{1}$}{S1} with point-like interactions} \label{sec:2}

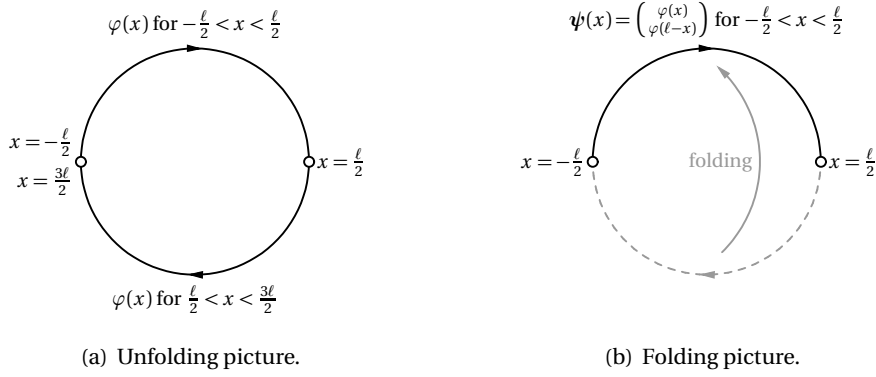
\begin{figure}[t]
\centering
\subfigure[Unfolding picture.]{
\xdefinecolor{rgb_000000}{rgb}{0,0,0}%
\xdefinecolor{rgb_ffffff}{rgb}{1,1,1}%
\setlength{\unitlength}{1cm}%
\begin{picture}(4.8,4.4)(0,0)%
\allinethickness{0.8pt}%
\path(3.96,2.2)(3.95704,2.29419)(3.94817,2.388)(3.93343,2.48107)
  (3.91287,2.57303)(3.88658,2.66353)(3.85466,2.75219)
  (3.81724,2.83867)(3.77446,2.92263)(3.72649,3.00374)
  (3.67353,3.08168)(3.61577,3.15614)(3.55345,3.22682)
  (3.48682,3.29345)(3.41614,3.35577)(3.34168,3.41353)
  (3.26374,3.46649)(3.18263,3.51446)(3.09867,3.55724)
  (3.01219,3.59466)(2.92353,3.62658)(2.83303,3.65287)
  (2.74107,3.67343)(2.648,3.68817)(2.55419,3.69704)(2.46,3.7)
  (2.36581,3.69704)(2.272,3.68817)(2.17893,3.67343)(2.08697,3.65287)
  (1.99647,3.62658)(1.90781,3.59466)(1.82133,3.55724)
  (1.73737,3.51446)(1.65626,3.46649)(1.57832,3.41353)
  (1.50386,3.35577)(1.43318,3.29345)(1.36655,3.22682)
  (1.30423,3.15614)(1.24647,3.08168)(1.19351,3.00374)
  (1.14554,2.92263)(1.10276,2.83867)(1.06534,2.75219)
  (1.03342,2.66353)(1.00713,2.57303)(0.986569,2.48107)
  (0.971828,2.388)(0.96296,2.29419)(0.96,2.2)(0.96296,2.10581)
  (0.971828,2.012)(0.986569,1.91893)(1.00713,1.82697)
  (1.03342,1.73647)(1.06534,1.64781)(1.10276,1.56133)
  (1.14554,1.47737)(1.19351,1.39626)(1.24647,1.31832)
  (1.30423,1.24386)(1.36655,1.17318)(1.43318,1.10655)
  (1.50386,1.04423)(1.57832,0.986475)(1.65626,0.933508)
  (1.73737,0.88554)(1.82133,0.842759)(1.90781,0.805335)
  (1.99647,0.773415)(2.08697,0.747125)(2.17893,0.726569)
  (2.272,0.711828)(2.36581,0.70296)(2.46,0.7)(2.55419,0.70296)
  (2.648,0.711828)(2.74107,0.726569)(2.83303,0.747125)
  (2.92353,0.773415)(3.01219,0.805335)(3.09867,0.842759)
  (3.18263,0.88554)(3.26374,0.933508)(3.34168,0.986475)
  (3.41614,1.04423)(3.48682,1.10655)(3.55345,1.17318)
  (3.61577,1.24386)(3.67353,1.31832)(3.72649,1.39626)
  (3.77446,1.47737)(3.81724,1.56133)(3.85466,1.64781)
  (3.88658,1.73647)(3.91287,1.82697)(3.93343,1.91893)(3.94817,2.012)
  (3.95704,2.10581)(3.96,2.2)
\put(3.96,2.2){\color{rgb_ffffff}$\allinethickness{0.0658987cm}\circle{0.0658987}$}%
\put(3.96,2.2){\color{rgb_000000}$\circle{0.131797}$}%
\put(0.96,2.2){\color{rgb_ffffff}$\allinethickness{0.0658987cm}\circle{0.0658987}$}%
\put(0.96,2.2){\color{rgb_000000}$\circle{0.131797}$}%
\path(2.46,3.7)(2.50712,3.69926)
\allinethickness{0.0140584cm}%
\path(2.36622,3.68038)(2.42623,3.68038)
\path(2.36644,3.69444)(2.48646,3.69444)
\path(2.36666,3.7085)(2.47249,3.7085)
\path(2.36688,3.72255)(2.4198,3.72255)
\path(2.49429,3.69627)(2.49429,3.70268)
\path(2.48146,3.69327)(2.48146,3.7061)
\path(2.46863,3.69028)(2.46863,3.70953)
\path(2.4558,3.68728)(2.4558,3.71295)
\path(2.44297,3.68429)(2.44297,3.71637)
\path(2.43014,3.6813)(2.43014,3.71979)
\path(2.41731,3.6783)(2.41731,3.72322)
\path(2.40448,3.67531)(2.40448,3.72664)
\path(2.39166,3.67231)(2.39166,3.73006)
\path(2.37883,3.66932)(2.37883,3.73348)
\allinethickness{0.8pt}%
\path(2.36655,3.70147)(2.366,3.66633)(2.50712,3.69926)
  (2.3671,3.73661)(2.36655,3.70147)
\path(2.46,0.7)(2.41288,0.70074)
\allinethickness{0.0140584cm}%
\path(2.55346,0.699081)(2.54117,0.73068)
\path(2.55292,0.664489)(2.52834,0.727686)
\path(2.53778,0.667423)(2.51552,0.724692)
\path(2.52217,0.671587)(2.50269,0.721698)
\path(2.50656,0.675752)(2.48986,0.718704)
\path(2.49095,0.679917)(2.47703,0.71571)
\path(2.47533,0.684081)(2.4642,0.712716)
\path(2.45972,0.688246)(2.45137,0.709722)
\path(2.44411,0.692411)(2.43854,0.706728)
\path(2.4285,0.696575)(2.42571,0.703734)
\path(2.49725,0.720429)(2.43289,0.695404)
\path(2.55393,0.729356)(2.45289,0.690069)
\path(2.55373,0.716163)(2.47289,0.684733)
\path(2.55352,0.70297)(2.49289,0.679397)
\path(2.55331,0.689777)(2.51289,0.674062)
\path(2.55311,0.676583)(2.5329,0.668726)
\allinethickness{0.8pt}%
\path(2.55345,0.698532)(2.554,0.733674)(2.41288,0.70074)
  (2.5529,0.66339)(2.55345,0.698532)
\put(4.06544,2.2){\makebox(0,0)[l]{\hbox{\color{rgb_000000}\scriptsize $x=\tfrac{\ell}{2}$}}}
\put(0.854562,2.27029){\makebox(0,0)[br]{\hbox{\color{rgb_000000}\scriptsize $x=-\tfrac{\ell}{2}$}}}
\put(0.854562,2.12971){\makebox(0,0)[tr]{\hbox{\color{rgb_000000}\scriptsize $x=\tfrac{3\ell}{2}$}}}
\put(2.46,3.84058){\makebox(0,0)[b]{\hbox{\color{rgb_000000}\scriptsize $\varphi(x)$ for $-\tfrac{\ell}{2} < x < \tfrac{\ell}{2}$}}}
\put(2.46,0.559416){\makebox(0,0)[t]{\hbox{\color{rgb_000000}\scriptsize $\varphi(x)$ for $\tfrac{\ell}{2} < x < \tfrac{3\ell}{2}$}}}
\end{picture}%
\label{fig:2a}}
\hspace*{5em}
\subfigure[Folding picture.]{
\xdefinecolor{rgb_000000}{rgb}{0,0,0}%
\xdefinecolor{rgb_999999}{rgb}{0.6,0.6,0.6}%
\xdefinecolor{rgb_ffffff}{rgb}{1,1,1}%
\setlength{\unitlength}{1cm}%
\begin{picture}(4.8,4.4)(0,0)%
\allinethickness{0.8pt}%
\path(3.96,2.2)(3.95704,2.29419)(3.94817,2.388)(3.93343,2.48107)
  (3.91287,2.57303)(3.88658,2.66353)(3.85466,2.75219)
  (3.81724,2.83867)(3.77446,2.92263)(3.72649,3.00374)
  (3.67353,3.08168)(3.61577,3.15614)(3.55345,3.22682)
  (3.48682,3.29345)(3.41614,3.35577)(3.34168,3.41353)
  (3.26374,3.46649)(3.18263,3.51446)(3.09867,3.55724)
  (3.01219,3.59466)(2.92353,3.62658)(2.83303,3.65287)
  (2.74107,3.67343)(2.648,3.68817)(2.55419,3.69704)(2.46,3.7)
  (2.36581,3.69704)(2.272,3.68817)(2.17893,3.67343)(2.08697,3.65287)
  (1.99647,3.62658)(1.90781,3.59466)(1.82133,3.55724)
  (1.73737,3.51446)(1.65626,3.46649)(1.57832,3.41353)
  (1.50386,3.35577)(1.43318,3.29345)(1.36655,3.22682)
  (1.30423,3.15614)(1.24647,3.08168)(1.19351,3.00374)
  (1.14554,2.92263)(1.10276,2.83867)(1.06534,2.75219)
  (1.03342,2.66353)(1.00713,2.57303)(0.986569,2.48107)
  (0.971828,2.388)(0.96296,2.29419)(0.96,2.2)
\color{rgb_999999}%
\path(0.96,2.2)(0.962957,2.153)
\path(0.968871,2.059)(0.971828,2.012)
\path(0.971828,2.012)(0.980652,1.96574)
\path(0.998301,1.87322)(1.00713,1.82697)
\path(1.00713,1.82697)(1.02168,1.78218)
\path(1.05078,1.6926)(1.06534,1.64781)
\path(1.06534,1.64781)(1.08539,1.6052)
\path(1.12549,1.51998)(1.14554,1.47737)
\path(1.14554,1.47737)(1.17077,1.43761)
\path(1.22124,1.35808)(1.24647,1.31832)
\path(1.24647,1.31832)(1.27649,1.28204)
\path(1.33653,1.20947)(1.36655,1.17318)
\path(1.36655,1.17318)(1.40088,1.14094)
\path(1.46953,1.07647)(1.50386,1.04423)
\path(1.50386,1.04423)(1.54196,1.01655)
\path(1.61816,0.961189)(1.65626,0.933508)
\path(1.65626,0.933508)(1.69753,0.910821)
\path(1.78006,0.865447)(1.82133,0.842759)
\path(1.82133,0.842759)(1.86512,0.825423)
\path(1.95269,0.790751)(1.99647,0.773415)
\path(1.99647,0.773415)(2.04209,0.761704)
\path(2.13331,0.738281)(2.17893,0.726569)
\path(2.17893,0.726569)(2.22565,0.720667)
\path(2.31909,0.708862)(2.36581,0.70296)
\path(2.36581,0.70296)(2.41291,0.70296)
\path(2.50709,0.70296)(2.55419,0.70296)
\path(2.55419,0.70296)(2.60091,0.708862)
\path(2.69435,0.720667)(2.74107,0.726569)
\path(2.74107,0.726569)(2.78669,0.738281)
\path(2.87791,0.761704)(2.92353,0.773415)
\path(2.92353,0.773415)(2.96731,0.790751)
\path(3.05488,0.825423)(3.09867,0.842759)
\path(3.09867,0.842759)(3.13994,0.865447)
\path(3.22247,0.910821)(3.26374,0.933508)
\path(3.26374,0.933508)(3.30184,0.961189)
\path(3.37804,1.01655)(3.41614,1.04423)
\path(3.41614,1.04423)(3.45047,1.07647)
\path(3.51912,1.14094)(3.55345,1.17318)
\path(3.55345,1.17318)(3.58347,1.20947)
\path(3.64351,1.28204)(3.67353,1.31832)
\path(3.67353,1.31832)(3.69876,1.35808)
\path(3.74923,1.43761)(3.77446,1.47737)
\path(3.77446,1.47737)(3.79451,1.51998)
\path(3.83461,1.6052)(3.85466,1.64781)
\path(3.85466,1.64781)(3.86922,1.6926)
\path(3.89832,1.78218)(3.91287,1.82697)
\path(3.91287,1.82697)(3.9217,1.87322)
\path(3.93935,1.96574)(3.94817,2.012)
\path(3.94817,2.012)(3.95113,2.059)
\path(3.95704,2.153)(3.96,2.2)
\path(2.46,0.7)(2.44822,0.700185)
\path(2.42466,0.700555)(2.41288,0.70074)
\allinethickness{0.0140584cm}%
\path(2.5002,0.677447)(2.55312,0.677447)
\path(2.44751,0.691504)(2.55334,0.691504)
\path(2.43354,0.70556)(2.55356,0.70556)
\path(2.49377,0.719617)(2.55378,0.719617)
\path(2.54117,0.666518)(2.54117,0.73068)
\path(2.52834,0.66994)(2.52834,0.727686)
\path(2.51552,0.673362)(2.51552,0.724692)
\path(2.50269,0.676785)(2.50269,0.721698)
\path(2.48986,0.680207)(2.48986,0.718704)
\path(2.47703,0.683629)(2.47703,0.71571)
\path(2.4642,0.687051)(2.4642,0.712716)
\path(2.45137,0.690474)(2.45137,0.709722)
\path(2.43854,0.693896)(2.43854,0.706728)
\path(2.42571,0.697318)(2.42571,0.703734)
\allinethickness{0.8pt}%
\path(2.55345,0.698532)(2.554,0.733674)(2.41288,0.70074)
  (2.5529,0.66339)(2.55345,0.698532)
\put(3.96,2.2){\color{rgb_ffffff}$\allinethickness{0.0658987cm}\circle{0.0658987}$}%
\put(3.96,2.2){\color{rgb_000000}$\circle{0.131797}$}%
\put(0.96,2.2){\color{rgb_ffffff}$\allinethickness{0.0658987cm}\circle{0.0658987}$}%
\put(0.96,2.2){\color{rgb_000000}$\circle{0.131797}$}%
\color{rgb_000000}%
\path(2.46,3.7)(2.50712,3.69926)
\allinethickness{0.0140584cm}%
\path(2.49429,3.69627)(2.4915,3.70342)
\path(2.48146,3.69327)(2.47589,3.70759)
\path(2.46863,3.69028)(2.46028,3.71175)
\path(2.4558,3.68728)(2.44467,3.71592)
\path(2.44297,3.68429)(2.42905,3.72008)
\path(2.43014,3.6813)(2.41344,3.72425)
\path(2.41731,3.6783)(2.39783,3.72841)
\path(2.40448,3.67531)(2.38222,3.73258)
\path(2.39166,3.67231)(2.36708,3.73551)
\path(2.37883,3.66932)(2.36654,3.70092)
\path(2.3871,3.73127)(2.36689,3.72342)
\path(2.40711,3.72594)(2.36669,3.71022)
\path(2.42711,3.7206)(2.36648,3.69703)
\path(2.44711,3.71527)(2.36627,3.68384)
\path(2.46711,3.70993)(2.36607,3.67064)
\path(2.48711,3.7046)(2.42275,3.67957)
\allinethickness{0.8pt}%
\path(2.36655,3.70147)(2.366,3.66633)(2.50712,3.69926)
  (2.3671,3.73661)(2.36655,3.70147)
\put(4.06544,2.2){\makebox(0,0)[l]{\hbox{\color{rgb_000000}\scriptsize $x=\tfrac{\ell}{2}$}}}
\put(0.854562,2.2){\makebox(0,0)[r]{\hbox{\color{rgb_000000}\scriptsize $x=-\tfrac{\ell}{2}$}}}
\put(2.46,3.84058){\makebox(0,0)[b]{\hbox{\color{rgb_000000}\scriptsize $\bm{\psi}(x)=\left(\begin{smallmatrix}\varphi(x)\\ \varphi(\ell-x)\end{smallmatrix}\right)$ for $-\tfrac{\ell}{2} < x < \tfrac{\ell}{2}$}}}
\color{rgb_999999}%
\path(2.66208,0.997918)(2.69925,1.03627)(2.73519,1.07577)
  (2.76987,1.11638)(2.80326,1.15806)(2.83533,1.20077)
  (2.86604,1.24446)(2.89536,1.28909)(2.92326,1.33463)
  (2.94972,1.38102)(2.97471,1.42822)(2.99821,1.47618)
  (3.02018,1.52485)(3.04062,1.57419)(3.0595,1.62415)(3.0768,1.67467)
  (3.0925,1.72572)(3.10659,1.77723)(3.11906,1.82916)(3.12989,1.88145)
  (3.13907,1.93406)(3.14659,1.98693)(3.15246,2.04002)
  (3.15665,2.09326)(3.15916,2.1466)(3.16,2.2)(3.15916,2.2534)
  (3.15665,2.30674)(3.15246,2.35998)(3.14659,2.41307)
  (3.13907,2.46594)(3.12989,2.51855)(3.11906,2.57084)
  (3.10659,2.62277)(3.0925,2.67428)(3.0768,2.72533)(3.0595,2.77585)
  (3.04062,2.82581)(3.02018,2.87515)(2.99821,2.92382)
  (2.97471,2.97178)(2.94972,3.01898)(2.92326,3.06537)
  (2.89536,3.11091)(2.86604,3.15554)(2.83533,3.19923)
  (2.80326,3.24194)(2.76987,3.28362)(2.73519,3.32423)
  (2.69925,3.36373)(2.66208,3.40208)
\path(2.66208,3.40208)(2.62373,3.43925)
\allinethickness{0.0140584cm}%
\path(2.63548,3.43245)(2.63068,3.42806)
\path(2.64722,3.42565)(2.63764,3.41687)
\path(2.65897,3.41885)(2.64459,3.40568)
\path(2.67071,3.41205)(2.65155,3.39449)
\path(2.68246,3.40525)(2.6585,3.3833)
\path(2.69421,3.39845)(2.66546,3.37212)
\path(2.70595,3.39165)(2.67241,3.36093)
\path(2.7177,3.38485)(2.67937,3.34974)
\path(2.72944,3.37806)(2.68632,3.33855)
\path(2.74119,3.37126)(2.69327,3.32736)
\path(2.66006,3.38079)(2.71001,3.32627)
\path(2.6276,3.43701)(2.7198,3.33636)
\path(2.66811,3.41356)(2.72958,3.34646)
\path(2.70863,3.3901)(2.73936,3.35656)
\allinethickness{0.8pt}%
\path(2.72469,3.34141)(2.74915,3.36665)(2.62373,3.43925)
  (2.70023,3.31617)(2.72469,3.34141)
\put(3.05456,2.2){\makebox(0,0)[r]{\hbox{\color{rgb_000000}\scriptsize \textcolor{black!40}{folding}}}}
\end{picture}%
\label{fig:2b}}
\caption{Folding the circle in half. Arrows indicate the direction of coordinates.}
\label{fig:2}
\end{figure}

To begin with, let us set up the model.
Let $x \in (-\ell/2, 3\ell/2)$ be the coordinate of circle of circumference $2\ell$, where $x = -\ell/2$ and $x = 3\ell/2$ are identified, and $\varphi(x)$ be a wavefunction on the circle.
Point-like interactions are located at $x = \pm \ell/2$.
For the following discussions it is convenient to introduce a two-component vector-valued wavefunction $\bm{\psi}(x)$ on the interval $(-\ell/2, \ell/2)$, whose upper- and lower-components are given by the wavefunctions on the upper- and lower-semicircles, respectively:
\begin{align}
\bm{\psi}(x)
:= 	\begin{pmatrix}
	\varphi(x) \\
	\varphi(\ell - x)
	\end{pmatrix},
	\quad -\frac{\ell}{2} < x < \frac{\ell}{2}. \label{eq:2.1}
\end{align}
As shown in figure \ref{fig:2}, introducing this vector-valued wavefunction effectively means to fold the circle in half.
Mathematically speaking, the vector-valued wavefunction \eqref{eq:2.1} means to consider the following Hilbert space:
\begin{align}
\mathcal{H}
&= 	L^{2}(-\tfrac{\ell}{2}, \tfrac{\ell}{2})
	\oplus L^{2}(\tfrac{\ell}{2}, \tfrac{3\ell}{2}) \nonumber\\
&\cong
	L^{2}(-\tfrac{\ell}{2}, \tfrac{\ell}{2}) \otimes \mathbb{C}^{2}. \label{eq:2.2}
\end{align}
In this way, quantum mechanics on the circle of circumference $2\ell$ with scalar-valued wavefunctions is always mapped into quantum mechanics on the interval of length $\ell$ with vector-valued wavefunctions.
Point-like interactions located at antipodal points of the circle are then translated into the problem of boundary conditions at the boundaries of the interval.
Borrowing the terminology of boundary conformal field theory \cite{Oshikawa:1996,Bachas:2001vj}, we call the original scalar quantum mechanics the unfolding picture and the vector quantum mechanics the folding picture; see figures \ref{fig:2a} and \ref{fig:2b}.

The time-independent Schr\"{o}dinger equation for a free particle in the folding picture is therefore given by the vector equation
\begin{align}
-\bm{\psi}^{\prime\prime} = E\bm{\psi}, \label{eq:2.3}
\end{align}
where prime (${}^{\prime}$) indicates the derivative with respect to $x$.
In this paper we are interested in point-like interactions that respect the unitarity of time evolution, or self-adjointness of the free Hamiltonian $H = \mathrm{diag}\,(-d^{2}/dx^{2}, -d^{2}/dx^{2}) = -d^{2}/dx^{2} \otimes 1_{2}$, where $1_{2}$ stands for the $2 \times 2$ unit matrix.
Such point-like interactions are all encoded into the probability current conservation conditions at the boundaries, $j(-\ell/2) = 0 = j(\ell/2)$, where $j(x) = -i(\bm{\psi}^{\dagger}(x)\bm{\psi}^{\prime}(x) - \bm{\psi}^{\prime\dagger}(x)\bm{\psi}(x))$ is the local probability current in the folding picture.
These conditions are quadratic in the boundary-valued wavefunctions; however, as shown by F\"{u}l\"{o}p and Tsutsui \cite{Fulop:1999pf}, they can be linearized and are known to enjoy the following $U(2)$ family of solutions at each boundary:
\begin{subequations}
\begin{align}
(1_{2} + U_{1})\bm{\psi}^{\prime} - iv(1_{2} - U_{1})\bm{\psi} = 0 \quad\text{at}\quad x = -\frac{\ell}{2}, \label{eq:2.4a}\\
(1_{2} + U_{2})\bm{\psi}^{\prime} - iv(1_{2} - U_{2})\bm{\psi} = 0 \quad\text{at}\quad x = +\frac{\ell}{2}, \label{eq:2.4b}
\end{align}
\end{subequations}
where $U_{1}$ and $U_{2}$ are $2 \times 2$ unitary matrices and $v$ is an arbitrary reference scale of length dimension $-1$ that needs to be introduced to adjust the length dimensions of $\bm{\psi}^{\prime}$ and $\bm{\psi}$.
Hence in general the parameter space of the model is $U(2) \times U(2)$.
In this paper, however, we focus on the following $SU(2) \subset U(2) \times U(2)$ subfamily of boundary conditions described by a single special unitary matrix $U = U_{1} = U_{2} \in SU(2)$:
\begin{align}
(1_{2} + U)\bm{\psi}^{\prime} - iv(1_{2} - U)\bm{\psi} = 0
\quad\text{at}\quad
x = \pm\frac{\ell}{2}. \label{eq:2.5}
\end{align}
We will see that, with this choice of boundary conditions, all the energy levels become doubly-degenerate and the ground states are given by zero-modes of certain first-order differential operators.
We shall then show that, when the unitary matrix $U$ becomes time-dependent and varies very slowly along a closed loop on the parameter space, geometric phases are given by the path-ordered exponentials of 't Hooft-Polyakov monopoles of $SU(2)$ Yang-Mills-Higgs theory.

To see this, let us first parameterize the special unitary matrix $U \in SU(2)$.
As is well-known, the parameter space of $SU(2)$ is 3-sphere, which becomes clear in the following parameterization:
\begin{align}
U
= 	\begin{pmatrix}
	x_{1} + ix_{2} 	& ix_{3} + x_{4} \\
	ix_{3} - x_{4} 	& x_{1} - ix_{2}
	\end{pmatrix}, \label{eq:2.6}
\end{align}
where $(x_{1}, x_{2}, x_{3}, x_{4}) \in \mathbb{R}^{4}$ satisfies the condition $x_{1}^{2} + x_{2}^{2} + x_{3}^{2} + x_{4}^{2} = 1$ and hence parameterizes the unit 3-sphere $S^{3}$.
For the following discussions, however, it is convenient to work in the following spectral decomposition parameterization:
\begin{align}
U = \mathrm{e}^{i\alpha}P_{+} + \mathrm{e}^{-i\alpha}P_{-}, \quad
P_{\pm} = \frac{1_{2} \pm Z}{2}, \label{eq:2.7}
\end{align}
where $\alpha \in [0, \pi]$ is an angle parameter and $P_{\pm}$ are projection operators.
$Z$ is a generic hermitian traceless unitary matrix that satisfies $Z = Z^{\dagger} = Z^{-1}$ and $Z^{2} = 1_{2}$.
Such $2 \times 2$ hermitian unitary matrix can be parameterized as follows:
\begin{align}
Z = \bm{n}\cdot\bm{\sigma}, \label{eq:2.8}
\end{align}
where $\bm{\sigma} = (\sigma_{1}, \sigma_{2}, \sigma_{3})$ is the vector of Pauli matrices and $\bm{n} = (n_{1}, n_{2}, n_{3})$ is a real unit 3-vector that fulfils the condition $n_{1}^{2} + n_{2}^{2} + n_{3}^{2} = 1$; that is, $\bm{n}$ parameterizes the unit 2-sphere.
It is wise to compare here these two different parameterizations \eqref{eq:2.6} and \eqref{eq:2.7}.
To this end, let us parameterize the unit 3-vector $\bm{n}$ into the following spherical coordinates:
\begin{align}
\bm{n} = (\sin\theta\cos\phi, \sin\theta\sin\phi, \cos\theta), \label{eq:2.9}
\end{align}
where $\theta \in [0, \pi]$ and $\phi \in [0, 2\pi)$.
Substituting \eqref{eq:2.8} into \eqref{eq:2.7} with the parameterization \eqref{eq:2.9} and then comparing the result with \eqref{eq:2.6} we see that the unit 4-vector $(x_{1}, x_{2}, x_{3}, x_{4})$ is given as follows:
\begin{align}
\begin{cases}
x_{1} = \cos\alpha, \\
x_{2} = \sin\alpha\cos\theta, \\
x_{3} = \sin\alpha\sin\theta\cos\phi, \\
x_{4} = \sin\alpha\sin\theta\sin\phi,
\end{cases} \label{eq:2.10}
\end{align}
which is the standard spherical coordinates for the unit 3-sphere.
An important point to note here is that the eigenphase $\alpha$ ranges from $0$ to $\pi$, not from $0$ to $2\pi$, which becomes crucial in section \ref{sec:3.1}.

Now, it is convenient to introduce orthonormal eigenvectors $\{\bm{e}_{+}, \bm{e}_{-}\}$ of $Z$ that satisfy $Z\bm{e}_{\pm} = \pm\bm{e}_{\pm}$.
Such eigenvectors satisfy the eigenvalue equations $U\bm{e}_{\pm} = \mathrm{e}^{\pm i\alpha}\bm{e}_{\pm}$, the orthonormality $\bm{e}_{\alpha}^{\dagger}\bm{e}_{\beta} = \delta_{\alpha\beta}$ and the completeness $\sum_{\alpha=\pm}\bm{e}_{\alpha}\bm{e}_{\alpha}^{\dagger} = 1_{2}$.
Notice that the projection operators can be written as $P_{\pm} = \bm{e}_{\pm}\bm{e}_{\pm}^{\dagger}$.
Since the set of eigenvectors $\{\bm{e}_{+}, \bm{e}_{-}\}$ provides the complete orthonormal basis of 2-dimensional vector space, any element $\bm{\psi}$ of the Hilbert space \eqref{eq:2.2} can be decomposed as follows:
\begin{align}
\bm{\psi}(x) = \psi_{+}(x)\bm{e}_{+} + \psi_{-}(x)\bm{e}_{-}, \label{eq:2.11}
\end{align}
where $\psi_{\pm}(x) = \bm{e}_{\pm}^{\dagger}\bm{\psi}(x)$.
In this paper we call the set of eigenvectors $\{\bm{e}_{+}, \bm{e}_{-}\}$ the basis and the coefficient functions $\{\psi_{+}, \psi_{-}\}$ the components.
Now it is easy to see that the boundary conditions \eqref{eq:2.5} reduce to the following Robin boundary conditions for the components:
\begin{align}
\psi_{+}^{\prime} - v(\alpha)\psi_{+} = 0
\quad\text{and}\quad
\psi_{-}^{\prime} + v(\alpha)\psi_{-} = 0
\quad\text{at}\quad
x = \pm\frac{\ell}{2}, \label{eq:2.12}
\end{align}
where $v(\alpha) := v\tan(\alpha/2)$.
Notice that the Schr\"{o}dinger equation \eqref{eq:2.3} boils down to the following two independent differential equations for the components:
\begin{align}
-\psi_{\pm}^{\prime\prime} = E\psi_{\pm}. \label{eq:2.13}
\end{align}

\subsection{Doubly-degenerate energy levels} \label{sec:2.1}
Now it is a straightforward exercise to solve the Schr\"{o}dinger equations \eqref{eq:2.13} with the Robin boundary conditions \eqref{eq:2.12}.
As the equation \eqref{eq:2.12} implies, the ground states are given by the zero-modes of the first-order differential operators $d/dx \mp v(\alpha)$.
A straightforward calculation shows that the normalized ground-state energy eigenfunctions $\bm{\psi}_{\pm,0} = \psi_{\pm,0}\bm{e}_{\pm}$ take the following exponential forms localized on the boundaries:
\begin{align}
\bm{\psi}_{\pm,0}(x)
= 	\sqrt{\frac{v(\alpha)}{\sinh(v(\alpha)\ell)}}
	\exp\left(\pm v(\alpha)x\right)\bm{e}_{\pm}. \label{eq:2.14}
\end{align}
The ground state energy is independent of the size of the interval $\ell$ and given by
\begin{align}
E_{0} = - v(\alpha)^{2}. \label{eq:2.15}
\end{align}
It is easy to see that excited states are also doubly degenerate.
Normalized energy eigenfunctions $\bm{\psi}_{\pm, n} = \psi_{\pm, n}\bm{e}_{\pm}$ take the following forms:
\begin{align}
\bm{\psi}_{\pm,n}(x)
= 	\sqrt{\frac{2}{\ell}\frac{1}{1+(v(\alpha)/k_{n})^{2}}}
	\left[
	\cos\left(k_{n}\left(x+\frac{\ell}{2}\right)\right)
	\pm
	\frac{v(\alpha)}{k_{n}}\sin\left(k_{n}\left(x+\frac{\ell}{2}\right)\right)
	\right]
	\bm{e}_{\pm},
	\quad
	n = 1,2,\cdots, \label{eq:2.16}
\end{align}
where wave numbers $k_{n}$ are given by $k_{n} = n\pi/\ell$.
The energy eigenvalues are independent of the boundary condition parameter $v(\alpha)$ and given by
\begin{align}
E_{n} = \left(\frac{n\pi}{\ell}\right)^{2},
\quad
n = 1,2,\cdots. \label{eq:2.17}
\end{align}
As we will see in the subsequent sections, the relative phase between $\bm{\psi}_{+,n}$ and $\bm{\psi}_{-,n}$ is chosen to satisfy the higher-derivative supersymmetry relations \eqref{eq:2.19} and is the most convenient choice for the computation of Berry's connections.

\subsection{Higher-derivative supersymmetry} \label{sec:2.2}
It would be reasonable to expect that there might be some underlying symmetry that ensures two-fold degeneracy.
In this section we shall show that the symmetry behind the doubly-degenerate energy levels is the so-called second-order derivative supersymmetry \cite{Andrianov:1993md,Andrianov:1994aj}, which is a nonlinear extension of $\mathscr{N} = 2$ quantum mechanical supersymmetry.
Just as in the case of ordinary $\mathscr{N} = 2$ supersymmetry, the second-order derivative supersymmetry algebra consists of four operators: the Hamiltonian $H$, which is the second-order derivative operator, the supercharge $Q^{+}$ and its adjoint $Q^{-}$, which are also the second-order derivative operators, and the fermion parity $(-1)^{F}$, which is the $\mathbb{Z}_{2}$ grading operator.
Below we shall construct these operators explicitly and discuss its algebraic structures briefly.

\begin{figure}[t]
\centerline{\input{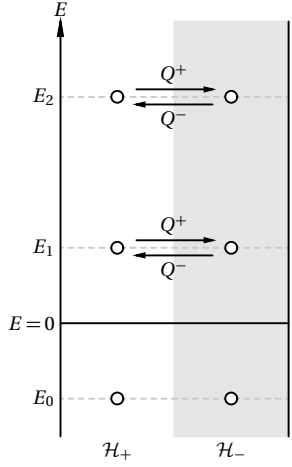}}
\caption{Schematic structure of doubly-degenerate energy levels. Arrows indicate the second-order derivative supersymmetry transformations between ``bosonic'' and ``fermionic'' states $|\bm{\psi}_{+,n}\rangle$ and $|\bm{\psi}_{-,n}\rangle$.}
\label{fig:3}
\end{figure}

To this end, let us first introduce the following first-order differential operators:
\begin{align}
A^{\pm} = \pm\frac{d}{dx} + v(\alpha), \label{eq:2.18}
\end{align}
whose zero-modes are the ground-state wavefunctions \eqref{eq:2.14}.
It is easy to check that the components $\psi_{\pm,n} = \bm{e}_{\pm}^{\dagger}\bm{\psi}_{\pm,n}$ of the excited states \eqref{eq:2.16} satisfy the following relations:
\begin{align}
A^{\mp}A^{\mp}\psi_{\pm,n} = \pm (E_{n} - E_{0})\psi_{\mp,n}. \label{eq:2.19}
\end{align}
Now we are in a position to introduce the second-order derivative supersymmetry algebra.
To this end, let us work in the basis in which $U$ becomes diagonal.
In this basis the energy eigenfunctions take the forms $\bm{\psi}_{+,n} = (\psi_{+,n}, 0)^{T}$ and $\bm{\psi}_{-,n} = (0, \psi_{-,n})^{T}$.
The set of operators $\{H, Q^{\pm}, (-1)^{F}\}$ is then given by the following standard forms \cite{Andrianov:1993md,Andrianov:1994aj}:
\begin{subequations}
\begin{align}
H
&= 	\begin{pmatrix}
	A^{+}A^{-} + E_{0} 	& 0 \\
	0 				& A^{-}A^{+} + E_{0}
	\end{pmatrix}, \label{eq:2.20a}\\
Q^{+}
&= 	\begin{pmatrix}
	0 		& 0 \\
	A^{-}A^{-} 	& 0
	\end{pmatrix}, \label{eq:2.20b}\\
Q^{-}
&= 	\begin{pmatrix}
	0 	& A^{+}A^{+} \\
	0 	& 0
	\end{pmatrix}, \label{eq:2.20c}\\
(-1)^{F}
&= 	\begin{pmatrix}
	1 	& 0 \\
	0 	& -1
	\end{pmatrix}, \label{eq:2.20d}
\end{align}
\end{subequations}
which act on the energy eigenfunctions as $H\bm{\psi}_{\pm,n} = E_{n}\bm{\psi}_{\pm,n}$, $Q^{\pm}\bm{\psi}_{\pm,n} = \pm (E_{n} - E_{0})\bm{\psi}_{\mp,n}$, $(-1)^{F}\bm{\psi}_{\pm,n} = \pm\bm{\psi}_{\pm,n}$, and satisfy the following relations of second-order derivative supersymmetry algebra:
\begin{subequations}
\begin{align}
&(Q^{\pm})^{2} = 0, \label{eq:2.21a}\\
&\left((-1)^{F}\right)^{2} = 1_{2}, \label{eq:2.21b}\\
&[H, Q^{\pm}] = [H, (-1)^{F}] = 0, \label{eq:2.21c}\\
&\{Q^{\pm}, (-1)^{F}\} = 0, \label{eq:2.21d}\\
&\{Q^{+}, Q^{-}\} = (H - E_{0})^{2}. \label{eq:2.21e}
\end{align}
\end{subequations}
Since these algebraic relations are invariant under any unitary transformation $\mathcal{O} \mapsto \Tilde{\mathcal{O}} = U\mathcal{O}U^{\dagger}$, where $\mathcal{O} \in \{H, Q^{\pm}, (-1)^{F}\}$ and $U \in U(2)$, the equations \eqref{eq:2.21a}--\eqref{eq:2.21e} hold true in any basis.
We note that, thanks to the fermion parity $(-1)^{F}$ which commutes with the Hamiltonian, the Hilbert space splits into two orthogonal subspaces:
\begin{align}
\mathcal{H} = \mathcal{H}_{+} \oplus \mathcal{H}_{-}, \label{eq:2.22}
\end{align}
where $\mathcal{H}_{\pm} = \{\bm{\psi} \in \mathcal{H}: (-1)^{F}\bm{\psi} = \pm\bm{\psi}\}$ are ``bosonic'' and ``fermionic'' subspaces, respectively.
Except for the ground states, elements of $\mathcal{H}_{+}$ and $\mathcal{H}_{-}$ are transformed into each other by the second-order derivative supersymmetry transformations $Q^{\pm}\bm{\psi}_{\pm,n} = \pm (E_{n} - E_{0})\bm{\psi}_{\mp,n}$, which is schematically illustrated in figure \ref{fig:3}.

\section{Geometric phase in the ground-state sector} \label{sec:3}
Let us now move on to the analysis of time-dependent situation where boundary condition parameters vary very slowly.
Suppose that the initial state $\bm{\psi}_{\text{in}}$ at time $t=0$ is in the subspace of $n$th excited states, $\bm{\psi}_{\text{in}} \in \mathcal{H}_{n} = \mathrm{span}\{\bm{\psi}_{+,n}, \bm{\psi}_{-,n}\}$.
At time $t=T$, transitions between different subspaces $\mathcal{H}_{n}$ and $\mathcal{H}_{m}$ ($n \neq m$) are suppressed by the factor $1/T$ such that the initial state $\bm{\psi}_{\text{in}} \in \mathcal{H}_{n}$ remains in the subspace $\mathcal{H}_{n}$ in the adiabatic limit $T \to \infty$.
Under an adiabatic time-evolution along a closed path $\gamma$ on the parameter space, the initial state $\bm{\psi}_{\text{in}}$ transforms into the final state $\bm{\psi}_{\text{out}} = \mathrm{e}^{-i\int_{0}^{T}\!\!dt\,E_{n}}W_{\gamma}(A^{(n)})\bm{\psi}_{\text{in}} \in \mathcal{H}_{n}$, where $\mathrm{e}^{-i\int_{0}^{T}\!\!dt\,E_{n}}$ is the $T$-dependent trivial dynamical phase and $W_{\gamma}(A^{(n)})$ is the $T$-independent nontrivial non-Abelian geometric phase given by the Wilson loop \cite{Wilczek:1984dh}
\begin{align}
W_{\gamma}(A^{(n)}) = \mathcal{P}\mathrm{exp}\left(i\oint_{\gamma}A^{(n)}\right). \label{eq:3.1}
\end{align}
Here $A^{(n)} = \left(\begin{smallmatrix}A^{(n)}_{++} & A^{(n)}_{+-}\\ A^{(n)}_{-+} & A^{(n)}_{--}\end{smallmatrix}\right)$ is the $2 \times 2$ hermitian matrix-valued one-form, or Berry's connection, for the $n$th excited sector and is given by the inner product
\begin{align}
A^{(n)}_{ab}
= 	i\langle\bm{\psi}_{a,n}|d|\bm{\psi}_{b,n}\rangle
:= 	i\int_{-\ell/2}^{\ell/2}\!\!\!dx\,
	\bm{\psi}_{a,n}^{\dagger}(x)d\bm{\psi}_{b,n}(x),
	\quad a,b \in \{+,-\}, \label{eq:3.2}
\end{align}
where $d$ stands for the exterior derivative on the space of boundary conditions $SU(2) \cong S^{3}$.
It should be noted that under the unitary change of the basis of the subspace $\mathcal{H}_{n} = \mathrm{span}\{\bm{\psi}_{+,n}, \bm{\psi}_{-,n}\}$,
\begin{align}
\bm{\psi}_{a,n} \mapsto \Tilde{\bm{\psi}}_{a,n} = \bm{\psi}_{b,n}g_{ba},
\quad g = (g_{ba}) \in SU(2), \label{eq:3.3}
\end{align}
the Berry connection \eqref{eq:3.2} indeed transforms as a connection:
\begin{align}
A^{(n)} \mapsto \Tilde{A}^{(n)} = g^{\dagger}A^{(n)}g + ig^{\dagger}dg. \label{eq:3.4}
\end{align}
In the subsequent sections we shall compute the Berry connections \eqref{eq:3.2} in two different gauges, the ``string'' and ``hedgehog'' gauges, and show that they are given by the BPS 't Hooft-Polyakov monopole for $n=0$ and non-BPS 't Hooft-Polyakov monopoles for $n \geq 1$.

\subsection{Berry's connection = BPS monopole} \label{sec:3.1}
Let us first focus on the ground-state sector.
Substituting the ground-state solutions \eqref{eq:2.14} into the equation \eqref{eq:3.2} we get
\begin{align}
A^{(0)}_{ab} = iK^{(0)}_{ab}\bm{e}_{a}^{\dagger}d\bm{e}_{b}, \label{eq:3.5}
\end{align}
where $K^{(0)}_{ab}$ is the overlapping integral between the components $\psi_{a,0}$ and $\psi_{b,0}$ given by
\begin{align}
K^{(0)}_{ab}
= 	\int_{-\ell/2}^{\ell/2}\!\!\!dx\,
	\psi_{a,0}^{\ast}(x)\psi_{b,0}(x). \label{eq:3.6}
\end{align}
A straightforward calculation gives
\begin{align}
K^{(0)}_{\pm\pm} = 1
\quad\text{and}\quad
K^{(0)}_{\pm\mp} = \frac{v(\alpha)\ell}{\sinh(v(\alpha)\ell)} =: K^{(0)}. \label{eq:3.7}
\end{align}
The Berry connection in the ground-state sector then takes the form
\begin{align}
A^{(0)}
= 	\begin{pmatrix}
	i\bm{e}_{+}^{\dagger}d\bm{e}_{+} 		& iK^{(0)}\bm{e}_{+}^{\dagger}d\bm{e}_{-} \\
	iK^{(0)}\bm{e}_{-}^{\dagger}d\bm{e}_{+} 	& i\bm{e}_{-}^{\dagger}d\bm{e}_{-}
	\end{pmatrix}. \label{eq:3.8}
\end{align}
Let us first compute this gauge field by parameterizing the orthonormal eigenvectors $\{\bm{e}_{+}, \bm{e}_{-}\}$.
In the spherical coordinates \eqref{eq:2.9} the unitary matrix $Z = \bm{n}\cdot\bm{\sigma}$ is parameterized as $Z = \left(\begin{smallmatrix}\cos\theta & \mathrm{e}^{-i\phi}\sin\theta \\ \mathrm{e}^{i\phi}\sin\theta & -\cos\theta\end{smallmatrix}\right)$.
The orthonormal eigenvectors of $Z$ are then chosen to be of the forms
\begin{align}
\bm{e}_{+}
= 	\begin{pmatrix}
	\cos\frac{\theta}{2} \\
	\mathrm{e}^{i\phi}\sin\frac{\theta}{2}
	\end{pmatrix},
\quad
\bm{e}_{-}
= 	\begin{pmatrix}
	-\mathrm{e}^{-i\phi}\sin\frac{\theta}{2} \\
	\cos\frac{\theta}{2}
	\end{pmatrix}. \label{eq:3.9}
\end{align}
Substituting these into the equation \eqref{eq:3.8} we get $A^{(0)} = A^{(0)}_{\theta}d\theta + A^{(0)}_{\phi}d\phi$, where
\begin{subequations}
\begin{align}
A^{(0)}_{\theta}
&= 	- K^{(0)}\sin\phi\frac{\sigma_{1}}{2}
	+ K^{(0)}\cos\phi\frac{\sigma_{2}}{2}, \label{eq:3.10a}\\
A^{(0)}_{\phi}
&= 	- K^{(0)}\sin\theta\cos\phi\frac{\sigma_{1}}{2}
	- K^{(0)}\sin\theta\sin\phi\frac{\sigma_{2}}{2}
	- (1 - \cos\theta)\frac{\sigma_{3}}{2}. \label{eq:3.10b}
\end{align}
\end{subequations}
Notice that $A^{(0)}_{\phi}$ is ill-defined at the south pole $\theta = \pi$ because the combination $1 - \cos\theta$ does not vanish at $\theta = \pi$.
In other words, the Berry connection suffers from the Dirac string singularity along the negative 3-axis.
This Dirac string singularity comes from the fact that the eigenvectors \eqref{eq:3.9} are not well-defined at $\theta = \pi$.
(Notice that the combinations $\mathrm{e}^{\pm i\phi}\sin(\theta/2)$ do not vanish at $\theta = \pi$.)
In fact, the eigenvectors $\{\bm{e}_{+}, \bm{e}_{-}\}$ cannot be globally well-defined over the whole 2-sphere in any way.
This Dirac string singularity, however, can be removed by singular gauge transformation \cite{Arafune:1974uy}.
Below we shall perform such singular gauge transformation and then show that the Berry connection \eqref{eq:3.8} indeed represents the BPS monopole.

To do this, let us move into the following gauge:
\begin{align}
g
= 	\begin{pmatrix}
	\bm{e}_{+}^{\dagger} \\
	\bm{e}_{-}^{\dagger}
	\end{pmatrix}. \label{eq:3.11}
\end{align}
Notice that this unitary matrix inherits the Dirac string singularity from the eigenvectors $\{\bm{e}_{+}, \bm{e}_{-}\}$ and hence is not globally well-defined over the whole parameter space.
In this gauge the Berry connection \eqref{eq:3.4} becomes the following simple form (see appendix \ref{appendix:A}):
\begin{align}
\Tilde{A}^{(0)} = \frac{i}{2}\left(1 - K^{(0)}\right)ZdZ. \label{eq:3.12}
\end{align}
Let us next parameterize the unit 3-vector $\bm{n}$ into the following ``hedgehog'' configuration:
\begin{align}
\bm{n} = \frac{\bm{r}}{r}, \label{eq:3.13}
\end{align}
where $\bm{r} = (x_{1}, x_{2}, x_{3}) \in \mathbb{R}^{3}$ is a real 3-vector and $r = \sqrt{x_{1}^{2} + x_{2}^{2} + x_{3}^{2}} \geq 0$ is its length.
With this parameterization the one-form $iZdZ$ becomes $iZdZ = \epsilon_{ijk}(x_{j}\sigma_{k}/r^{2})dx_{i}$.
Though $r$ can be arbitrary from the viewpoint of the parameterization of $\bm{n}$, in the rest of this paper we fix the length $r$ to satisfy the following relation:
\begin{align}
r = \ell\tan\left(\frac{\alpha}{2}\right). \label{eq:3.14}
\end{align}
Notice that $\alpha$ runs from $0$ to $\pi$ such that $\ell\tan(\alpha/2)$ is non-negative.
Under this identification the Berry connection $\Tilde{A}^{(0)} = \Tilde{A}^{(0)}_{i}dx_{i}$ is cast into the following manifestly spherically symmetric form:
\begin{align}
\Tilde{A}^{(0)}_{i}
= 	\epsilon_{ijk}\frac{x_{j}}{r^{2}}\frac{\sigma_{k}}{2}
	\left(1 - \frac{vr}{\sinh(vr)}\right), \label{eq:3.15}
\end{align}
where we have used $v(\alpha)\ell = v\ell\tan(\alpha/2) = vr$.
This is nothing but the well-known BPS monopole solution of $SU(2)$ Yang-Mills-Higgs theory \cite{Prasad:1975kr}, where $v$ plays the role of Higgs vacuum expectation value.
Notice that the Dirac string singularity disappears in this gauge.
In fact, there is no singularity in \eqref{eq:3.15}, which is consistent with the fact that there is no additional spectral degeneracy or level crossings in the whole parameter region.

\subsection{Matrix elements of position operator = Higgs field} \label{sec:3.2}
In ref.~\cite{Sonner:2008be} Sonner and Tong constructed a quantum mechanical model for a spin-1/2 on $S^{2}$ where Berry's connection and the matrix elements of the operator $\cos\Hat{\theta}$ become the BPS solutions for gauge and Higgs fields of $SU(2)$ Yang-Mills-Higgs theory, respectively.
Motivated by their results, in this section we would like to construct a quantum mechanical counterpart of Higgs field in our model.

To this end, let us consider the following matrix elements of position operator $\Hat{x}$ in the ground-state sector:\footnote{This is the matrix elements of position operator in the folding picture. In the unfolding picture it is give by
\begin{align}
\Phi^{(0)}_{ab}
&= 	\frac{v}{\ell}\int_{-\ell/2}^{3\ell/2}\!\!\!dx\,
	\varphi_{a,0}^{\ast}(x)
	\left[
	\theta(\tfrac{\ell}{2} - x)x + \theta(x - \tfrac{\ell}{2})(\ell - x)
	\right]
	\varphi_{b,0}(x), \nonumber
\end{align}
where $\theta(x)$ is the Heaviside step function.}
\begin{align}
\Phi^{(0)}_{ab}
= 	\frac{v}{\ell}\langle\bm{\psi}_{a,0}|\Hat{x}|\bm{\psi}_{b,0}\rangle
:= 	\frac{v}{\ell}\int_{-\ell/2}^{\ell/2}\!\!\!dx\,
	\bm{\psi}_{a,0}^{\dagger}(x)x\bm{\psi}_{b,0}(x),
	\quad
	a,b \in \{+,-\}, \label{eq:3.16}
\end{align}
where the overall factor $v/\ell$ is introduced for later convenience.
It should be emphasized that the gauge transformation \eqref{eq:3.3} acts on the $2 \times 2$ hermitian matrix $\Phi^{(0)} = \left(\begin{smallmatrix}\Phi^{(0)}_{++} & \Phi^{(0)}_{+-}\\ \Phi^{(0)}_{-+} & \Phi^{(0)}_{--}\end{smallmatrix}\right)$ as the adjoint action
\begin{align}
\Phi^{(0)} \mapsto \Tilde{\Phi}^{(0)} = g^{\dagger}\Phi^{(0)}g. \label{eq:3.17}
\end{align}
A straightforward calculation shows that the matrix elements take the forms
\begin{align}
\Phi^{(0)}_{ab} = vH^{(0)}_{a}\delta_{ab}, \label{eq:3.18}
\end{align}
where
\begin{align}
H^{(0)}_{\pm}
&= 	\frac{1}{\ell}\int_{-\ell/2}^{\ell/2}\!\!\!dx\,
	\psi_{\pm,0}^{\ast}(x)x\psi_{\pm,0}(x)
= 	\pm\frac{1}{2}\left(\coth(v(\alpha)\ell) - \frac{1}{v(\alpha)\ell}\right)
=: 	\pm H^{(0)}. \label{eq:3.19}
\end{align}
Hence the hermitian matrix $\Phi^{(0)}$ is given by the following diagonal matrix:
\begin{align}
\Phi^{(0)}
= 	\begin{pmatrix}
	vH^{(0)} 	& 0 \\
	0 		& -vH^{(0)}
	\end{pmatrix}
= 	vH^{(0)}\sigma_{3}. \label{eq:3.20}
\end{align}
This is, again, not so clear that it gives the BPS solution.
To see this, let us move into the gauge \eqref{eq:3.11} just as we did in the previous section.
As shown in appendix \ref{appendix:A}, in this gauge the hermitian matrix \eqref{eq:3.17} becomes
\begin{align}
\Tilde{\Phi}^{(0)} = vH^{(0)}Z. \label{eq:3.21}
\end{align}
With the same parameterization as \eqref{eq:3.13} and \eqref{eq:3.14}, $\Tilde{\Phi}^{(0)}$ takes the following form:
\begin{align}
\Tilde{\Phi}^{(0)}
= 	v\frac{x_{i}}{r}\frac{\sigma_{i}}{2}
	\left(\coth(vr) - \frac{1}{vr}\right), \label{eq:3.22}
\end{align}
which is exactly the same form as the BPS solution for the Higgs field in $SU(2)$ Yang-Mills-Higgs theory \cite{Prasad:1975kr}.

\section{Geometric phase in the excited-state sector} \label{sec:4}
Let us next move on to the analysis of Berry's connections in the excited-state sectors.
Substituting the solutions \eqref{eq:2.16} into \eqref{eq:3.2} we find
\begin{align}
A^{(n)}
&= 	\begin{pmatrix}
	i\bm{e}_{+}^{\dagger}d\bm{e}_{+} 		& iK^{(n)}\bm{e}_{+}^{\dagger}d\bm{e}_{-} \\
	iK^{(n)}\bm{e}_{-}^{\dagger}d\bm{e}_{+} 	& i\bm{e}_{-}^{\dagger}d\bm{e}_{-}
	\end{pmatrix}. \label{eq:4.1}
\end{align}
Here $K^{(n)}$ is given by the following overlapping integral:
\begin{align}
K^{(n)}
= 	\int_{-\ell/2}^{\ell/2}\!\!\!dx\,
	\psi_{\pm,n}^{\ast}(x)\psi_{\mp,n}(x)
= 	\frac{1 - (vr/n\pi)^{2}}{1 + (vr/n\pi)^{2}},
	\quad
	n = 1,2,\cdots, \label{eq:4.2}
\end{align}
where $r$ is defined in \eqref{eq:3.14}.
Notice that the Berry connection \eqref{eq:4.1} takes the same form as that in the ground-state sector \eqref{eq:3.8} except for the explicit form of $K^{(n)}$.
Hence the gauge transformation \eqref{eq:3.4} with \eqref{eq:3.11} yields the same result (see appendix \ref{appendix:A})
\begin{align}
\Tilde{A}^{(n)} = \frac{i}{2}\left(1 - K^{(n)}\right)ZdZ. \label{eq:4.3}
\end{align}
By making use of the ``hedgehog'' parameterization \eqref{eq:3.13} we see that Berry's connection $\Tilde{A}^{(n)} = \Tilde{A}^{(n)}_{i}dx_{i}$ takes the following form:
\begin{align}
\Tilde{A}^{(n)}_{i}
= 	\epsilon_{ijk}\frac{x_{j}}{r^{2}}\frac{\sigma_{k}}{2}
	\left(1 - \frac{1 - (vr/n\pi)^{2}}{1 + (vr/n\pi)^{2}}\right),
	\quad
	n = 1,2,\cdots. \label{eq:4.4}
\end{align}
This can be interpreted as a 't Hooft-Polyakov monopole of $SU(2)$ Yang-Mills-Higgs theory.
To see this, we note that $K^{(n)}$ has the following asymptotic behaviors:
\begin{subequations}
\begin{align}
K^{(n)}
&\to 	+1+O(r^{+2}) \quad\text{as}\quad r \to 0, \label{eq:4.5a}\\
K^{(n)}
&\to 	-1+O(r^{-2}) \quad\text{as}\quad r \to \infty. \label{eq:4.5b}
\end{align}
\end{subequations}
Thus the Berry connection $\Tilde{A}^{(n)} = (i/2)(1 - K^{(n)})ZdZ$ vanishes at $r = 0$ and becomes pure gauge $iZdZ$ at $r = \infty$, which is the desired properties of 't Hooft-Polyakov monopole \cite{'tHooft:1974qc,Polyakov:1974ek}.

\section{Conclusions} \label{sec:5}
In this paper we have studied Berry's connections in quantum mechanical system for a free spinless particle on a circle with two point-like interactions.
We first showed that, for an $SU(2) \subset U(2) \times U(2)$ subfamily of point-like interactions, all the energy levels become doubly-degenerate thanks to the hidden higher-derivative supersymmetry.
We then showed that in this system Berry's connection $A^{(n)} = A^{(n)}_{i}dx_{i}$ for the $n$th excited sector in the ``hedgehog'' gauge always takes the form of 't Hooft-Polyakov ansatz \cite{'tHooft:1974qc,Polyakov:1974ek}
\begin{align}
A^{(n)}_{i} = \epsilon_{ijk}\frac{x_{j}}{r^{2}}\frac{\sigma_{k}}{2}\left(1 - K^{(n)}(r)\right),
\quad
n = 0,1,\cdots, \label{eq:5.1}
\end{align}
where $K^{(n)} \to 1$ as $r \to 0$ and $K^{(n)} \to -1$ as $r \to \infty$.
In particular, it was shown that the 't Hooft-Polyakov monopole \eqref{eq:5.1} becomes BPS for the ground-state sector $n = 0$.
Motivated by the work of Sonner and Tong \cite{Sonner:2008be}, we also explored a quantum mechanical counterpart of Higgs field in our model and showed that the matrix elements of position operator in the ground-state sector gives the BPS solution of Higgs field in $SU(2)$ Yang-Mills-Higgs theory.

It should be noted that our construction of BPS $SU(2)$ solutions in the ground-state sector is essentially equivalent to Nahm's construction \cite{Nahm:1979yw} (see also, e.g., chapter 4 of \cite{Weinberg:2006rq} for review).
Since Nahm's construction is applicable to any other gauge groups, it might be interesting to investigate quantum mechanical systems where Berry's connection and matrix elements of some operator give the BPS solutions for other gauge groups.

\subsection*{Acknowledgement}
The author would like to thank the Czech Technical University in Prague, where part of this work was carried out under the ESF grant CZ.1.07/2.3.00/30.0034.

\appendix
\titleformat{\section}[block]{\filright\bfseries\mathversion{bold}}{Appendix \thesection.}{0.5em}{}[\titlerule]
\section{Singular gauge transformations} \label{appendix:A}
In this section we derive the results \eqref{eq:3.12}, \eqref{eq:3.21} and \eqref{eq:4.3} by performing the singular gauge transformations that remove the Dirac string singularities.

Let us first focus on the Berry connection.
In the singular gauge \eqref{eq:3.11}, Berry's connections \eqref{eq:3.8} and \eqref{eq:4.1} transform as follows:
\begin{align}
\Tilde{A}^{(n)}
&= 	\begin{pmatrix}
	\bm{e}_{+} 	& \bm{e}_{-}
	\end{pmatrix}
	\begin{pmatrix}
	i\bm{e}_{+}^{\dagger}d\bm{e}_{+} 		& iK^{(n)}\bm{e}_{+}^{\dagger}d\bm{e}_{-} \\
	iK^{(n)}\bm{e}_{-}^{\dagger}d\bm{e}_{+} 	& i\bm{e}_{-}^{\dagger}d\bm{e}_{-}
	\end{pmatrix}
	\begin{pmatrix}
	\bm{e}_{+}^{\dagger} \\
	\bm{e}_{-}^{\dagger}
	\end{pmatrix}
	+ i
	\begin{pmatrix}
	\bm{e}_{+} 	& \bm{e}_{-}
	\end{pmatrix}
	\begin{pmatrix}
	d\bm{e}_{+}^{\dagger} \\
	d\bm{e}_{-}^{\dagger}
	\end{pmatrix} \nonumber\\
&= 	i(\bm{e}_{+}\bm{e}_{+}^{\dagger}d\bm{e}_{+}\bm{e}_{+}^{\dagger} + \bm{e}_{+}d\bm{e}_{+}^{\dagger})
	+ i(\bm{e}_{-}\bm{e}_{-}^{\dagger}d\bm{e}_{-}\bm{e}_{-}^{\dagger} + \bm{e}_{-}d\bm{e}_{-}^{\dagger}) \nonumber\\
& 	\hspace{1em}
	+ iK^{(n)}\bm{e}_{+}\bm{e}_{+}^{\dagger}d\bm{e}_{-}\bm{e}_{-}^{\dagger}
	+ iK^{(n)}\bm{e}_{-}\bm{e}_{-}^{\dagger}d\bm{e}_{+}\bm{e}_{+}^{\dagger}.  \label{eq:A.1}
\end{align}
This expression looks cumbersome but it can be simplified by using two different expressions of the projection operators $P_{\pm} = \bm{e}_{\pm}\bm{e}_{\pm}^{\dagger} = (1_{2} \pm Z)/2$.
To do this, we first note that the following identity holds:
\begin{align}
P_{a}dP_{b}
&= 	\bm{e}_{a}\bm{e}_{a}^{\dagger}
	\left(
	d\bm{e}_{b}\bm{e}_{b}^{\dagger}
	+ \bm{e}_{b}d\bm{e}_{b}^{\dagger}
	\right)
= 	\bm{e}_{a}\bm{e}_{a}^{\dagger}d\bm{e}_{b}\bm{e}_{b}^{\dagger}
	+ \delta_{ab}\bm{e}_{a}d\bm{e}_{a}^{\dagger}. \label{eq:A.2}
\end{align}
Plugging back into \eqref{eq:A.1} we get
\begin{align}
\Tilde{A}^{(n)}
&= 	iP_{+}dP_{+} + iP_{-}dP_{-} + iK^{(n)}P_{+}dP_{-} + iK^{(n)}P_{-}dP_{+} \nonumber\\
&= 	i\sum_{a,b=\pm}K^{(n)}_{ab}P_{a}dP_{b}, \label{eq:A.3}
\end{align}
where $K^{(n)}_{\pm\pm} = 1$ and $K^{(n)}_{\pm\mp} = K^{(n)}$ with $K^{(n)}$ given in \eqref{eq:3.7} and \eqref{eq:4.2}.
Substituting another parameterization of the projection operators $P_{\pm} = (1_{2} \pm Z)/2$ into \eqref{eq:A.3}, we get the following simple result:
\begin{align}
\Tilde{A}^{(n)}
&= 	\frac{i}{2}\left(1 - K^{(n)}\right)ZdZ. \label{eq:A.4}
\end{align}

Let us next consider the hermitian matrix \eqref{eq:3.17}.
In the gauge choice \eqref{eq:3.11} $\Tilde{\Phi}^{(0)}$ is given by
\begin{align}
\Tilde{\Phi}^{(0)}
&= 	\begin{pmatrix}
	\bm{e}_{+} 	& \bm{e}_{-}
	\end{pmatrix}
	\begin{pmatrix}
	vH^{(0)} 	& 0 \\
	0 		& -vH^{(0)}
	\end{pmatrix}
	\begin{pmatrix}
	\bm{e}_{+}^{\dagger} \\
	\bm{e}_{-}^{\dagger}
	\end{pmatrix} \nonumber\\
&= 	vH^{(0)}(\bm{e}_{+}\bm{e}_{+}^{\dagger} - \bm{e}_{-}\bm{e}_{-}^{\dagger}) \nonumber\\
&= 	vH^{(0)}(P_{+} - P_{-}). \label{eq:A.5}
\end{align}
Noting that $Z = P_{+} - P_{-}$ we obtain the final result
\begin{align}
\Tilde{\Phi}^{(0)}
&= 	vH^{(0)}Z. \label{eq:A.6}
\end{align}

\bibliographystyle{utphys}
\bibliography{Bibliography}

\end{document}